\documentclass[final]{article}

\usepackage[dvips]{epsfig,psfrag}
\usepackage{float}
\usepackage{amsmath}
\usepackage{amssymb}
\usepackage{amsxtra}
\usepackage{enumerate}
\usepackage{boxedminipage}
\usepackage{longtable}

\newcommand{\R}{I\!\!R}

\title{An Introduction to Using Software Tools
for Automatic Differentiation (Revision A)
\thanks{ANL/MCS-TM-254. 
This work was supported by the Mathematical, Information, and
Computational Sciences Division subprogram of the Office of Advanced
Scientific Computing Research, U.S. Department of Energy, under Contract
W-31-109-ENG-38.
}
}

\author{Uwe Naumann and Andrea Walther \thanks{Mathematics and Computer Science Division, 
Argonne National Laboratory, Argonne, IL,
{\tt naumann@mcs.anl.gov} and 
Institute for Scientific
Computing, Technical University Dresden, Germany, 
{\tt awalther@math.tu-dresden.de}}.} 

\begin{document}
\pagenumbering{roman}
\setcounter{page}{1}
\thispagestyle{empty}
\begin{center}
\vspace*{-1in}
Argonne National Laboratory \\
9700 South Cass Avenue\\
Argonne, IL 60439

\vspace{.2in}
\rule{2in}{.01in}\\ [1ex]
ANL/MCS-TM-254, Revision A \\
\rule{2in}{.01in}

\vspace{1.5in}
{\large\bf An Introduction to Using Software Tools\\  
for Automatic Differentiation\footnote{This
work was supported by the Mathematical,
Information, and Computational Sciences Division subprogram of the
Office of Advanced Scientific Computing Research, Office of Science, U.S. Department of
Energy, under Contract W-31-109-Eng-38.
}
}

\vspace{.2in}
by \\ [3ex]

{\large\it Uwe Naumann and Andrea Walther\footnote{Institute for Scientific
Computing, Technical University Dresden, Germany}
}\\ [1ex]
{\tt naumann@mcs.anl.gov awalther@math.tu-dresden.de}

\thispagestyle{empty}

\vspace{1in}
Mathematics and Computer Science Division

\bigskip

Technical Memorandum No. 254, Revision A

\vspace{1in}
July 2003
\end{center}

\newpage
\pagenumbering{roman}
\setcounter{page}{2}
Argonne National Laboratory, with facilities in the states of Illinois and Idaho, is 
owned by the United States Government and operated by The University of Chicago under the provisions of a contract with the Department of Energy.

\vspace{1in}

\begin{center}
DISCLAIMER
\end{center}

This report was prepared as an account of work sponsored by an agency of the United States Government. Neither the United States Government nor any agency thereof, nor The University of Chicago, nor any of their employees or officers, makes any warranty, express or implied, or assumes any legal liability or responsibility for the accuracy, completeness, or usefulness of any information, apparatus, product, or process disclosed, or represents that its use would not infringe privately owned rights. Reference herein to any specific commercial product, process, or service by trade name, trademark, manufacturer, or otherwise, does not necessarily constitute or imply its endorsement, recommendation, or favoring by the United States Government or any agency thereof. The views and opinions of document authors expressed herein do not necessarily state or reflect those of the United States Government or any agency thereof.

\vspace{1in}

\noindent
Available electronically at http://www.doe.gov/bridge

\bigskip
\noindent
Available for a processing fee to U.S. Dept. of Energy and its contractors,
in paper, from:

\begin{center}
\begin{flushleft}
U.S. Department of Energy\\
Office of Scientific and Techdnical Information\\
P.O. Box 62\\
Oak Ridge, TN 37831-0062\\
phone: (865) 576-8401\\
fax: (865) 576-5728\\
email: reports@adonis.osti.gov
\end{flushleft}
\end{center}
  \pagestyle{plain}
\newpage
\tableofcontents
\newpage
\clearpage

\pagenumbering{arabic}
\setcounter{page}{1}

\begin{center}
{\large\bf An Introduction to Using Software Tools \\
for Automatic Differentiation} \\ [2ex]
by \\ [2ex]
{\large Uwe Naumann and Andrea Walther}
\end{center}
\addcontentsline{toc}{section}{Abstract}
\begin{abstract}
We give a gentle introduction to using various software tools for automatic
differentiation (AD). Ready-to-use examples are discussed, and links to
further information are presented. Our target audience includes all those
who are looking for a straightforward way to get started using the available AD
technology. The document is dynamic in the sense that its
content will be updated as the AD software evolves.
\end{abstract}

\section{Intention and References}

This document explains how to use the following software tools for automatic
differentiation (AD)
to generate first-order derivative code for a small example problem.
\begin{itemize}
\item ADIFOR 2.0 Revision D ({\tt http://www.mcs.anl.gov/adifor}) 
\item ADIC 1.1 ({\tt http://www.mcs.anl.gov/adic})
\item ADOL-C 1.8 ({\tt http://www.math.tu-dresden.de/wir/project/adolc})
\item TAPENADE 1.0-alpha ({\tt http://www-sop.inria.fr/tropics})
\end{itemize}
The document does not describe the AD algorithms used by these software tools. 
Various publications cover the theoretical foundations
of the algorithms referenced in our discussion. For an introduction
and more advanced issues we 
point the reader to \cite{Gri00}. 
Three workshops have been dedicated to AD, and the proceedings
\cite{BBCG96, CFG+02, CoGr91} contain numerous references to other
AD-related
papers that have appeared in scientific journals and proceedings of 
international conferences. A large variety of successful applications
of AD software to real-world applications are discussed there as well.

\section{Getting Started}

For information on how to obtain, install, and use the AD
tools discussed in this document, visit the Web sites listed
in Section 1.
The examples in this document are discussed in the context of the computer
environment found on a laptop computer, specifically an Intel Pentium
system running Mandrake Linux Version 8.1. ADIFOR, ADIC, and TAPENADE are 
installed in the directory {\tt /home/uwe/ADTOOLS}. The ADOL-C library and
include files are assumed to be in the same directory as the source
files.

ADIFOR, ADIC, and TAPENADE are available on the Web and can be
accessed online by using the corresponding Web servers. Visit
\begin{itemize} 
\item {\tt http://www.mcs.anl.gov/autodiff/adiforserver} for ADIFOR,
\item {\tt http://www.mcs.anl.gov/autodiff/adicserver} for ADIC, and
\item {\tt http://tapenade.inria.fr:8080/tapenade/index.jsp} for TAPENADE.
\end{itemize}

All source files presented in this document can be downloaded from 
\begin{center}
{\tt http://www-unix.mcs.anl.gov/\~\!naumann/ad\_tools.html}. 
\end{center}
The makefiles have to be adapted to the user's environment.

\section{Explosion Equation}

We have selected a variant of the Bratu problem \cite{ACM91} 
to illustrate the use of source transformation AD tools.
The code in  
Appendix \ref{app:1} models
the thermal explosion of solid fuels,
which can be described by the system of differential equations
$$
x''(\tau)+s \cdot e^{\frac{x(\tau)}{1+t \cdot x(\tau)}}=0,
$$
where $\tau \in (-1,1)$ and $x(-1)=x(1)=0.$
The problem has been discretized by using step size $h$ as
$$ 
F_i=x_{i-1}-2x_i +x_{i+1} +h^2[f_{i-1}+10 f_i +f_{i+1}] / 12
$$
for $i=1,\ldots,10000,$ with $x_0=x_{10001}=0$ and
$f_i=s \exp(x_i/(1+t x_i)).$
Of interest are the derivatives of the component functions $F_i$
with respect to the current state $x_i$ as well as the parameters
$s$ and $t.$ 

We consider the following problems:
\begin{enumerate}
\item ADIFOR
\begin{enumerate}
\item Use the forward vector mode to compute the Jacobian of 
$F$ with respect to $x$ and the parameters
$s$ and $t$. In our implementation, $F$ is represented by the variable
{\tt f}, $x$ is represented by {\tt x}, and
$s$ and $t$ are combined into the parameter vector {\tt prm}.
All three program variables are declared as arrays of double-precision 
floating-point numbers, namely,
\begin{itemize}
\item {\tt double precision x(7), prm(2), f(dim)} in Fortran and
\item {\tt double x[7], prm[2], f[7]} in C.
\end{itemize}
(See Section~\ref{ssec:ADIFOR}.)
\item Use Curtis-Powell-Reid \cite{CPR74} seeding to compute the
compressed Jacobian in forward vector mode
(see Section~\ref{sssec:CPR}).
\item Use the SparsLinC library to compute the sparse Jacobian
by sparse forward mode
(see Section~\ref{sssec:SparsLinC}).
\end{enumerate}
\item ADOL-C
\begin{enumerate}
\item Use one of the easy-to-use drivers to compute the Jacobian of $F$
with respect to $x$ and the parameters $s$ and $t$.
\item Compute  the sparsity structure of the Jacobian.
\item Use the low-level routines for forward and reverse mode to compute the 
Jacobian of $F$ with respect to $x$ and the parameters $s$ and $t$.
\end{enumerate}
(See Section~\ref{ssec:ADOLC}.)
\item ADIC
\begin{enumerate}
\item Compute the Jacobian of $F$ with respect to 
$x$ in forward vector mode
(see Section~\ref{ssec:ADIC}).
\end{enumerate}
\item TAPENADE
\begin{enumerate}
\item Compute the Jacobian of $F$ with respect to
both $x$ and the parameters $s$ and $t$ in reverse mode
(see Section~\ref{ssec:TAPENADE}).
\end{enumerate}
\end{enumerate}

\subsection{ADIFOR 2.0} \label{ssec:ADIFOR}

ADIFOR generates a differentiated version of the subroutine
{\tt expl(...)} with the following header.
\begin{verbatim}
      subroutine g_expl(g_p_, dim, parmax, x, g_x, ldg_x, prm, g_prm, 
     + ldg_prm, f, g_f, ldg_f)
        integer dim, parmax
        double precision x(dim), prm(parmax)
        double precision f(dim)

        integer g_pmax_
        parameter (g_pmax_ = 9)
        integer g_p_, ldg_x, ldg_prm, ldg_f
        double precision g_x(ldg_x, dim), g_prm(ldg_prm, parmax), 
     +    g_f(ldg_f, dim)
\end{verbatim}
The parameter
{\tt g\_pmax\_} is the maximal number of directional derivatives
that can be computed. Its value is set by using {\tt AD\_PMAX}
in {\tt explosion.adf}. 
The parameter
is required because of the lack of dynamic memory
allocation in Fortran~77. The derivative components of all scalar
program variables are allocated as vectors of length 
{\tt g\_pmax\_}. For example, in order to compute the Jacobian matrix of 
{\tt f} with respect to {\tt x}, the value of 
{\tt g\_pmax\_} must be at least equal to 7, so that the derivative
components of {\tt x} can accommodate the identity
in $\R^7.$

The actual number of directional derivatives 
{\tt g\_p\_} must be less than or equal to 
{\tt g\_pmax\_}. The parameter
{\tt g\_p\_} determines the upper bound for the loops that compute
the values of the derivative components.

To compute the Jacobian of {\tt f} with respect to both {\tt x} and
{\tt prm}, we set 
{\tt g\_pmax\_} to be greater than or equal to the sum of the numbers
of elements in both vectors, that is, $9=7+2={\dim}+{\tt parmax}$. 
Consequently,
{\tt g\_x} contains the first {\tt dim} columns of the  seed matrix
and {\tt g\_prm} its last {\tt parmax} columns. The argument
{\tt g\_f} is the transpose of the 
$({\tt dim} \times {\tt ldg\_f})$-Jacobian, where 
${\tt ldg\_f}$ must be initialized to 9. All this is done
in the driver routine, which is shown in Appendix~\ref{app:ADF1driver}.

The ``{\tt dense:}" section of the makefile in Appendix~\ref{app:ADFmake} 
builds an executable that produces the following result. \\
\begin{center}
\begin{minipage}[c]{9cm}
\begin{verbatim}
-1.88  1.01  0.    0.    0.    0.    0.   0.21 -0.48
 1.01 -1.87  1.01  0.    0.    0.    0.   0.39 -1.78
 0.    1.01 -1.87  1.01  0.    0.    0.   0.48 -2.69
 0.    0.    1.01 -1.87  1.01  0.    0.   0.55 -3.49
 0.    0.    0.    1.01 -1.87  1.01  0.   0.48 -2.69
 0.    0.    0.    0.    1.01 -1.87  1.01 0.39 -1.78
 0.    0.    0.    0.    0.    1.01 -1.88 0.21 -0.48
\end{verbatim}
\end{minipage}
\end{center} $\;$ \\
\\
The output has been formatted for better readability.
It shows the full $7 \times 9$ Jacobian evaluated at the
argument: \\
\begin{verbatim}
      x(1) = 1.72
      x(2) = 3.45
      x(3) = 4.16
      x(4) = 4.87
      x(5) = 4.16
      x(6) = 3.45
      x(7) = 1.72

      prm(1) = 1.3
      prm(2) = 0.245828
\end{verbatim}

\subsubsection{Compressed Jacobian -- Seeding}
\label{sssec:CPR}

The full Jacobian computed in the preceding section is sparse,
and its sparsity pattern can be visualized as follows.
$$
\begin{matrix}
*&*&&&&&&*&*\\
*&*&*&&&&&*&*\\
&*&*&*&&&&*&*\\
&&*&*&*&&&*&*\\
&&&*&*&*&&*&*\\
&&&&*&*&*&*&*\\
&&&&&*&*&*&*\\
\end{matrix}.
$$
Here, $*$ stands for a nonzero entry, and blanks represent {\em structural}
zero entries in the Jacobian. In other words, no dependence exists between
the corresponding dependent and independent variables.

Curtis-Powell-Reid (CPR) \cite{CPR74} seeding is based on the idea that certain 
columns of the Jacobian can be merged to share storage. 
For example, column 1 and column 7
could share one column, thus resulting in a compressed version
of the Jacobian. This implies that the sparsity pattern must be known
in advance in order to exploit matrix compression techniques.
Recall that the derivative code
generated by ADIFOR always loops over the derivative components of all
{\em active} variables.\footnote{An active variable {\tt w} can be 
characterized as follows. At some point in the program 
(1) the value of {\tt w} depends on the value of some independent 
variables and
(2) there is some dependent variable whose value depends on {\tt w}.
Variables that are not active are referred to as passive.}
Many of them are equal to zero, leading to
predictably trivial multiplications that one would like to avoid.
Therefore, instead of computing the Jacobian as a Jacobian times identity matrix
product, one could try to compute a compressed Jacobian using a
seed matrix with fewer columns than the identity. Since the number of independent
variables is often very large, the size of the seed matrix can become
much smaller, leading to a decreased complexity of the forward vector mode
Jacobian computation.

In CPR seeding, one considers the column incidence graph of the Jacobian to try 
to determine a minimal vertex coloring. Whenever two vertices share the same
color, the corresponding columns can be stored in the same column of the
compressed Jacobian. In our example, the column incidence graph has the
following structure.
\begin{center}
\epsfig{file=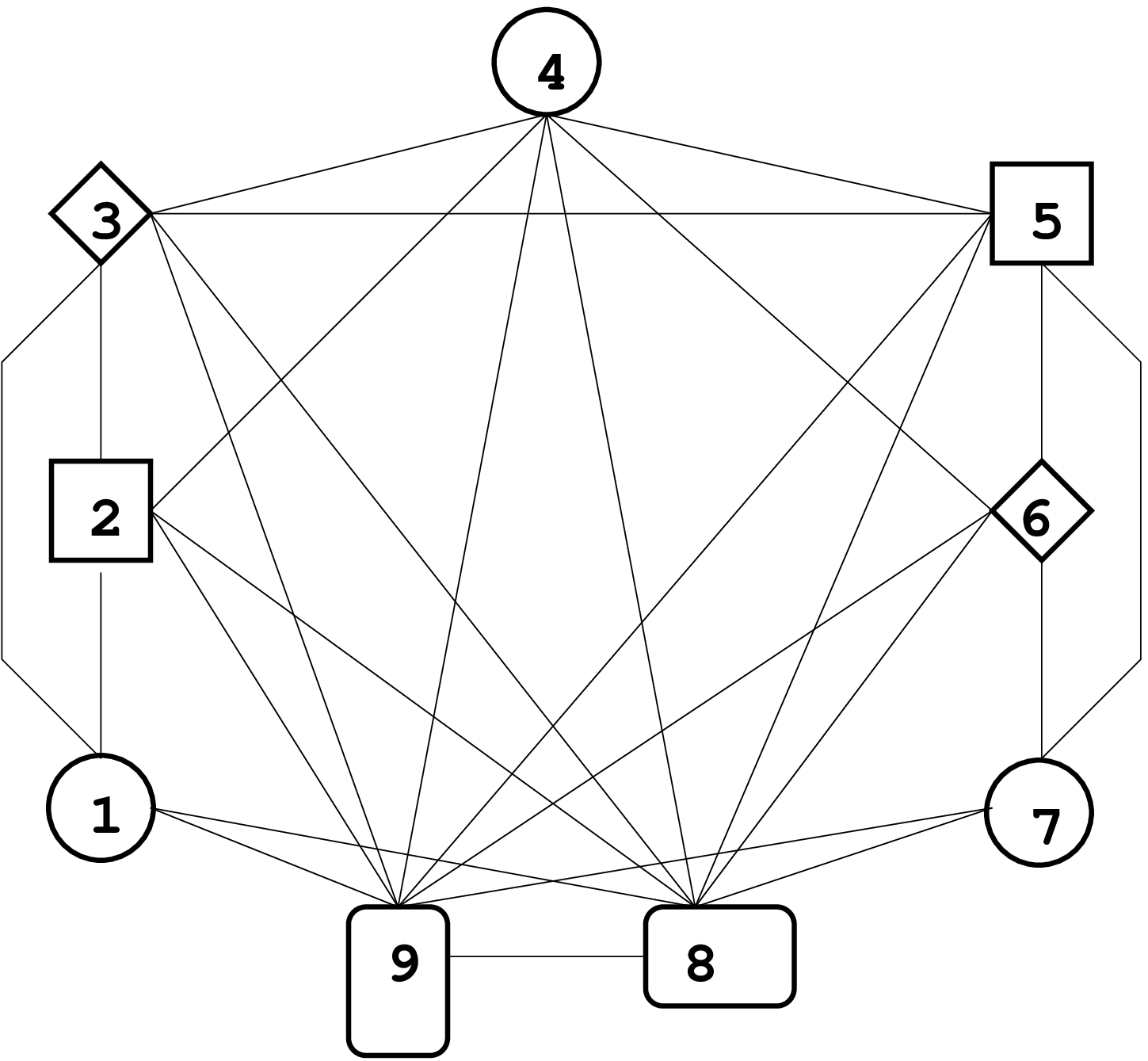,width=6cm}
\end{center}
Unfortunately, since the vertex coloring problem is known to be NP-complete
in general, the use of heuristics is essential. 
The coloring in the
example graph has been found ``by inspection." Different colors
are represented by different vertex shapes.

The number $\nu$ of different colors used determines the number of columns
in the CPR seed matrix. Its rows are Cartesian basis vectors in $\R^\nu$.
Whenever two vertices share the same color, the corresponding rows in the seed
matrix contain the same Cartesian basis vector. This leads to the following
seed matrix for our example.
$$
\begin{matrix}
1&0&0&0&0\\
0&1&0&0&0\\
0&0&1&0&0\\
1&0&0&0&0\\
0&1&0&0&0\\
0&0&1&0&0\\
1&0&0&0&0\\
0&0&0&1&0\\
0&0&0&0&1\\
\end{matrix}
$$
The ``{\tt compressed:}" section of the makefile in 
Appendix~\ref{app:ADFmake} 
builds an executable that produces 
output similar to the following. \\
\begin{center}
\begin{minipage}[c]{7cm}
\begin{verbatim}
-1.88  1.01  0.    0.21 -0.48
 1.01 -1.87  1.01  0.39 -1.78
 1.01  1.01 -1.87  0.48 -2.69
-1.87  1.01  1.01  0.55 -3.49
 1.01 -1.87  1.01  0.48 -2.69
 1.01  1.01 -1.87  0.39 -1.78
-1.88  0.    1.01  0.21 -0.48
\end{verbatim}
\end{minipage}
\end{center}
$\;$\\
It shows the compressed $7 \times 5$ Jacobian evaluated at the
current argument. The reconstruction of the original Jacobian
is a simple substitution process as described in 
\cite[Chapter 7]{Gri00}.

\subsubsection{Sparse Forward Mode -- SparsLinC}
\label{sssec:SparsLinC}

We compute the Jacobian of {\tt f} with respect to {\tt x}.
The sparsity pattern of the Jacobian need not be known a priori.
Derivative components are sparse vectors of (index, value) pairs.
The computational overhead of sparse vector arithmetic results from
the index calculations. Structural sparsity of the extended Jacobian
\cite[Chapter 2]{Gri00}
is exploited by avoiding trivial multiplications by zero. The decision
about
when to apply runtime sparsity methods depends on the problem
structure.

The use of SparsLinC with our example
can be described by the following steps:
\begin{enumerate}
\item Use the file {\tt explosion.cmp} without change.
\item Add the entry {\tt AD\_FLAVOR=sparse} to {\tt explosion.adf}. 
\item Write driver program as shown in Appendix~\ref{app:sparslinc}.
\item Generate {\tt g\_explosion.f} by calling
{\tt Adifor2.1 AD\_SCRIPT=explosion.adf},
and copy it from {\tt output\_files} to the current directory.
\item Compile {\tt g\_explosion.f} and the driver program.
\item Link with {\tt ReqADIntrinsics-Linux86.o},
{\tt libADIntrinsics-Linux86.a}, and
\newline
{\tt libSparsLinC-Linux86.a}.
\item Run the executable.
\end{enumerate}

Let us have a closer look at the driver program.
The differentiated subroutine
\begin{center}
{\tt g\_expl(dim,parmax,x,g\_x,prm,f,g\_f)}
\end{center}
is generated in {\tt g\_explosion.f}, where 
{\tt g\_x} and {\tt f\_x} are integer arrays of
dimension {\tt dim}
containing pointers to the corresponding sparse derivative objects.
Seeding {\tt g\_x} is performed in the driver by
\begin{verbatim}
      do 10 i=1,dim
        g_x(i)=0
        CALL DSPSD(g_x(i),i,1.d0,1)
 10   continue
\end{verbatim}
Notice that {\tt g\_x(i)} must be initialized properly before
calling {\tt DSPSD}. The call initializes the sparse derivative object
pointed at by {\tt g\_x(i)} as ({\tt i,1.d0}). Consequently, after executing 
this loop, {\tt g\_x} contains the sparse identity in $\R^{\tt dim}.$

The sparse derivative components are extracted by
\begin{verbatim}
      do 30 i=1,dim
        g_f(i)=0
        CALL DSPXSQ(indvec,valvec,dim,g_f(i),outlen,info)
 30   continue
\end{verbatim}
where {\tt indvec} is the index vector and {\tt valvec} the corresponding
value vector making up the sparse representation of the derivative
object pointed at by {\tt g\_f(i)}. The parameter {\tt outlen} is the number of
nonzero entries in the {\tt i}{\it th} row of the Jacobian. Consequently,
the {\tt outlen} first entries of 
{\tt indvec} and {\tt valvec} define the nonzero entries 
in the {\tt i}{\it th} row of the Jacobian as index value pairs
({\tt indvec(j), valvec(j)}). 

The ``{\tt sparse:}" section of the makefile in Appendix~\ref{app:ADFmake} 
builds an executable that produces 
output similar to the following. \\
\begin{center}
\begin{minipage}[c]{7cm}
\begin{verbatim}
(1, -1.88) (2,  1.01)
(1,  1.01) (2, -1.87) (3,  1.01)
(2,  1.01) (3, -1.87) (4,  1.01)
(3,  1.01) (4, -1.87) (4,  1.01)
(4,  1.01) (5, -1.87) (6,  1.01)
(5,  1.01) (6, -1.87) (7,  1.01)
(6,  1.01) (7, -1.88)
\end{verbatim}
\end{minipage}
\end{center} $\;$ \\
\\
The sparse Jacobian is given by sparse row vectors whose nonzero
entries are represented by (index, value) pairs. The reconstruction
of the full Jacobian is straightforward.
Refer to \cite{ADIFORman} for further information on how to use
SparsLinC.

\subsection{ADIC 1.1}
\label{ssec:ADIC}

All files involved in using ADIC with the C version of 
our example problem are shown in Appendix~\ref{app:ADIC}.
The Jacobian of {\tt F} with respect to {\tt x} is computed in forward
vector mode. ADIC uses the derived data type {\tt DERIV\_TYPE}
to associate derivative components with active variables.
The data type {\tt InactiveDouble} is used to indicate 
that some floating-point variable is not active. 
The call of {\tt ad\_AD\_Init(dim)} causes the derivative code to use
only the first {\tt dim} elements of the derivative components. The latter
are vectors of length {\tt GRAD\_MAX} (see Appendix~\ref{app:ADIC}).
The vector {\tt x} is declared to contain the {\tt dim} independent
variables by calling
\begin{verbatim}
      ad_AD_SetIndepArray(x,dim);
      ad_AD_SetIndepDone();
\end{verbatim}
The function value components of {\tt DERIV\_TYPE} variables
are accessed by means of
the macro {\tt DERIV\_val}. 

ADIC generates a differentiated version of the subroutine {\tt explosion}
and names it {\tt ad\_explosion}. The argument list remains unchanged
because all AD-related information is encapsulated in the new data type
{\tt DERIV\_TYPE}. If the function returns a 
{\tt double}, however, the differentiated function becomes a procedure,
and the returned value becomes the first argument.
The derivative components of scalar variables of
this type can be extracted by calling the routine {\tt ad\_AD\_ExtractGrad}.
The actual function value can be extracted with {\tt ad\_AD\_ExtractVal}.
In the example we call 
{\tt ad\_AD\_ExtractGrad(jac,F[i])}
inside a loop over the {\tt dim} elements of the vector of dependent
variables {\tt F}. The auxiliary variable {\tt jac} is declared as a passive
vector of size {\tt dim} and contains the {\tt i}{\it th} row of the
Jacobian.

A makefile is provided to build the executable {\tt explosion.ad} in order to
compute the first seven columns of the Jacobian shown in 
Section~\ref{ssec:ADIFOR}.

\subsection{ADOL-C} \label{ssec:ADOLC}
The AD-tool ADOL-C is based on operator overloading. Using this
technique, one can log for each operation during the program execution
the operator and the variables that are involved. Hence, one obtains a new  
internal representation of the function evaluation. Based on the
generated execution log, ADOL-C computes the desired derivatives.

To apply
ADOL-C  for derivative calculations, one first has to modify the
evaluation program to record the internal representation called tape. This
modification starts with including header-file(s) that introduce the
new data types and functions. Here, the easiest way is to simply
include {\sf adolc.h}. Second, one has to define the part of
the program for which one wants to compute the derivatives. From now
on, this part is called the active section. The statement {\sf
  trace\_on(tag,{\it keep});}  determines the beginning of the active
section, the statement  {\sf trace\_off({\it file});} the end of the
active section.  The parameter {\sf tag} identifies the function to be
differentiated. Hence, several function representations can be kept at
the same time. 
Because of the shortness of this introduction, the optional parameters {\sf keep} and
{\sf file} are not explained here but are explained in the documentation
\cite{ADOL-C}. Third, 
one has to change the types of the independents to {\sf
adouble}s and mark them as independents using the overloaded operator
$<<=$. Similarly, one must mark the dependents using the overloaded operator
$>>=$. Finally, all variables that lie on the way from the independents
to the dependents must be declared as {\sf adouble}s. This step
includes the generation of a new function {\sf explosion\_ad}, which
contains the same source code as before but in the interface the
{\sf double}-variables have to be changed to {\sf adouble}-variables.
The required
modification of the original source code is illustrated by 
Appendix~\ref{app:adolc}.
\subsubsection{Jacobian Calculation with the Easy-to-Use Drivers}
After the tapes are generated during the execution of
the active section, the required derivative objects can be
calculated. 
For that purpose ADOL-C provides a variety of easy-to-use
drivers:  {\sf gradient(..)}, {\sf jacobian(..)}, {\sf
hessian(..)}, {\sf vec\_jac(..)} computing the product vector times Jacobian,
\newline
{\sf jac\_vec(..)} computing the product  Jacobian times vector, and so on.
Moreover, ADOL-C supplies routines evaluating the Taylor coefficient vectors
and their Jacobians with respect to the current state vector of
solution curves defined by ordinary differential equations. 
Furthermore, there are drivers for derivative tensors and for the
differentiation of implicit and inverse functions. 

If one wants to compute the Jacobian of $F$ with respect to $x$ and
the parameters $s$ and $t$, one has to declare a variable storing the
derivative information

\hspace*{2cm}{\sf double **J = myalloc(dim,dim+parmax);}

\noindent
where myalloc is provided by ADOL-C to allocate a two-dimensional
array. Then, the statement 

\hspace*{2cm}{\sf jacobian(tag,dim,dim+parmax,v,J);}

\noindent
causes the computation of the Jacobian of the function representation
contained in the tape with the number {\sf tag} at the point {\sf v}.
For a consistency check, the second and third parameter determine the number
of dependents and independents, respectively. Hence, only the two
statements given above have to be inserted after {\sf trace\_off()}
in order to compute the Jacobian of $F$.
\subsubsection{Calculation of Sparsity Pattern}
ADOL-C provides for the computation of sparsity patterns the
driver 

\hspace*{2cm}{\sf jac\_pat(tag,dim,dim+parmax,v,rb,cb,Jsp,option);}

\noindent
If one sets {\sf rb} and {\sf cb} to {\sf NULL}, 
{\sf jac\_pat} computes the sparsity structure of the complete Jacobian
at the point where the tape was generated and stores it in the {\sf
unsigned int}-array {\sf Jsp}. The corresponding statements that have to
be added to the source code are shown in Appendix D. If
a certain block structure of the Jacobian is known, the {\sf
unsigned int}-vectors {\sf rb} and {\sf cb} can be applied to describe
a compressed form of the independent and dependent variables.
\subsubsection{Forward and Reverse Mode Using Low-Level Routines}
Arbitrarily high-order derivatives can be calculated by using the
low-level functions of ADOL-C for the forward and reverse mode of AD.
These routines are explained in detail in a short reference available
from the ADOL-C Web page. In this article, only the computation of first-order
derivatives is sketched. If one wants the full Jacobian, it is
preferable to use vector modes of AD. For that purpose, ADOL-C
provides the drivers {\sf fov\_forward(...)} and {\sf
fov\_reverse(...)}. Here, {\sf fov} stands for {\sf f}irst-{\sf o}rder
{\sf v}ector. The other acronyms of the low-level routines have a
corresponding meaning. The statement

\hspace*{2cm}{\sf fov\_forward(tag,dim,dim+parmax,p,v,X,Fp,Y);}

\noindent
computes the derivative object {\sf Y} $= F'(v)${\sf X} for {\sf X}
$\in \R^{{\sf dim+parmax} \times p}$. The statement 

\hspace*{2cm}{\sf fov\_reverse(tag,dim,dim+parmax,q,U,Z);}

\noindent
computes the derivative object {\sf Z} $=$  {\sf U}$F'(v)$ for {\sf U}
$\in \R^{q \times {\sf dim}}$. To prepare this reverse sweep,
one has to call an appropriate {\sf forward} routines, the choice of
which is described in detail by the short reference mentioned above.
The code segments required for computing the full Jacobian with the
low-level routines are contained in Appendix D. 

\subsection{TAPENADE}
\label{ssec:TAPENADE}

Our last example uses the alpha version of TAPENADE~1.0 to illustrate
the use of reverse-mode AD for computing the
Jacobian of {\tt f} with respect to {\tt x} and {\tt prm}. All files
needed to run the example are described in 
Appendix~\ref{app:TAPENADE}.  The
makefile in Appendix~\ref{app:TAPENADEmake} can be used to generate the
executable {\tt explosion.ad}. Running the latter results in the computation
of the full Jacobian
as in Section~\ref{ssec:ADIFOR} but this time using the reverse mode of
AD.

The following command-line parameters are used to call {\tt tapenade}:
\begin{itemize}
\item ``{\tt -head}": The name of the head routine
\item ``{\tt -vars}": The names of the independent variables
\item ``{\tt -cl}": Switch indicating that reverse, or cotangent linear,
mode is used
\end{itemize}
It generates derivative code in {\tt explcl.f}, which must be compiled
together with the driver program shown in Appendix~\ref{app:TAPENADEdriver}
and linked with the routines for storing and restoring the values of the 
tape as described in \cite[Chapter 2]{Gri00}.

The current version of TAPENADE does not provide the reverse vector mode. 
Hence, 
the Jacobian must be computed as a sequence of Jacobian transposed
times vector products. This computation is done in the driver program by successively 
initializing the derivative components {\tt g\_f} of {\tt f} as
the Cartesian basis vectors in $\R^{\tt dim}.$ Thus, the Jacobian is
accumulated row by row at a computational complexity proportional
to the number of dependent variables. This feature becomes particularly
interesting in the case of large, single gradients. 
Refer to \cite{Gri00} for further details on reverse-mode 
AD.

\newpage

\appendix

\section{Original Code} \label{app:1}

\begin{verbatim}
      subroutine bratu(dim,parmax,x,prm,F)
      integer    dim    , parmax
C     independent variables
      double precision x(dim), prm(parmax)
C     dependent variables
      double precision F(dim)
C
      integer i      
      double precision h

      h = 2.0/(dim+1)
      F(1) = -2*x(1)+h*h*prm(1)/12.0*(1+10*exp(x(1)/(1.0+prm(2)*x(1)))) 
      F(2) = x(1)+h*h*prm(1)/12.0*exp(x(1)/(1.0+prm(2)*x(1))) 

      do 1 i=2,dim-1
        F(i-1) = F(i-1)+x(i)+h*h*prm(1)/12.0*exp(x(i)/(1.0+prm(2)*x(i))) 
        F(i) = F(i)-2*x(i)+h*h*prm(1)/1.2*exp(x(i)/(1.0+prm(2)*x(i))) 
        F(i+1) = x(i)+h*h*prm(1)/12.0*exp(x(i)/(1.0+prm(2)*x(i))) 
    1 continue     

      F(dim-1) = F(dim-1)+x(dim)+h*h*prm(1)/12.0*exp(x(dim)/(1.0
     *           +prm(2)*x(dim))) 
      F(dim) = F(dim)-2*x(dim) 
      F(dim) = F(dim)+h*h*prm(1)/12.0*(1+10*exp(x(dim)/(1.0
     *         +prm(2)*x(dim)))) 
      end
\end{verbatim}

\section{ADIFOR 2.0}

The following files can be downloaded from
\begin{center}
{\tt http://www-unix.mcs.anl.gov/\~\!naumann/ad\_tools.html}
\end{center}

\begin{verbatim}
explosion.adf                 
explosion.cmp                 
explosion.driver.compressed.f  
explosion.driver.f         
explosion.driver.sparse.f  
explosion.f
explosion.sparse.adf
makefile
\end{verbatim}

\subsection{explosion.cmp}

The composition file lists the names of all files containing subroutines
subject to differentiation. In our simple example there is just 
{\tt explosion.f}.

\subsection{makefile} \label{app:ADFmake}

The {\tt makefile} can be used for computing the full, compressed, and
sparse Jacobians using ADIFOR~2.0. It also ensures a proper cleanup
of all files that are generated automatically during this process.
\begin{verbatim}
AD_LIB=/home/uwe/ADTOOLS/ADIFOR2/ADIFOR2.0D.lib

dense:	
	Adifor2.1 AD_SCRIPT=explosion.adf
	cp output_files/g_explosion.f .
	g77 -g -c g_explosion.f explosion.driver.f
	g77 -g -o explosion.ad.dense -L$(AD_LIB)/lib *.o \
	$(AD_LIB)/lib/ReqADIntrinsics-Linux86.o \
	-lADIntrinsics-Linux86 

compressed:	
	Adifor2.1 AD_SCRIPT=explosion.adf
	cp output_files/g_explosion.f .
	g77 -g -c g_explosion.f explosion.driver.compressed.f
	g77 -g -o explosion.ad.compressed -L$(AD_LIB)/lib *.o \
	$(AD_LIB)/lib/ReqADIntrinsics-Linux86.o \
	-lADIntrinsics-Linux86 

sparse: 
	Adifor2.1 AD_SCRIPT=explosion.sparse.adf
	cp output_files/g_explosion.f .
	g77 -g -c g_explosion.f explosion.driver.sparse.f
	g77 -g -o explosion.ad.sparse -L$(AD_LIB)/lib *.o \
	$(AD_LIB)/lib/ReqADIntrinsics-Linux86.o \
	-lADIntrinsics-Linux86 \
	-lSparsLinC-Linux86

clean:	
	rm -fr output_files
	rm -fr AD_cache
	rm *.o
	rm g_*
	rm explosion.ad.*
	rm *\~
\end{verbatim}

\subsection{Jacobian} \label{app:expl:dense}

\subsubsection{explosion.adf} 
The Jacobian of the output variable {\tt f} with respect to the two
input variables {\tt x} and {\tt prm} of the top-level routine {\tt expl}
is computed. 
The parameter {\tt AD\_PMAX} must be set to the total number of
independent variables, that is, 9\footnote{This is because the Jacobian is computed
by forward vector mode with a seed matrix that is equal to the identity
in $\R^9.$}.

\begin{verbatim}
AD_PROG = explosion.cmp
AD_TOP  = expl
AD_PMAX = 9
AD_IVARS= x prm
AD_DVARS= f
\end{verbatim}

\subsubsection{explosion.driver.f} \label{app:ADF1driver}
\begin{verbatim}
      program main
      implicit none
C 
C     Example: Explosion Equation
C     Driver for computing Jacobian
C
      integer dim, parmax, n
      parameter (dim=7, parmax=2, n=9)  
      integer i,j
C     independent variables   
      double precision x(dim), prm(parmax)
C     derivative components of independent variables   
      double precision g_x(n,dim)
      double precision g_prm(n,parmax)
C     dependent variables
      double precision f(dim)
C     derivative components of dependent variables
      double precision g_f(n,dim)

C     Initialization of input variables
      x(1) = 1.72  
      x(2) = 3.45  
      x(3) = 4.16  
      x(4) = 4.87  
      x(5) = 4.16  
      x(6) = 3.45 
      x(7) = 1.72 
      prm(1) = 1.3  
      prm(2) = 0.245828 

C     Seeding (identity)
      do 20 i=1,n
        do 10 j=1,parmax
          if (i.eq.j+dim) then
            g_prm(i,j)=1.0
          else
            g_prm(i,j)=0.0
          endif
 10     continue
 20   continue

      do 40 i=1,n
        do 30 j=1,dim
          if (i.eq.j) then
            g_x(i,j)=1.0
          else
            g_x(i,j)=0.0
          endif
 30     continue
 40   continue

C     call differentiated subroutine
      call g_expl(n,dim,parmax,x,g_x,n,prm,g_prm,n,f,g_f,n);
     
C     print Jacobian
      do 60 i=1,dim
        do 50 j=1,n
          print*, "f'(", i, ",", j, ")=",g_f(j,i)
 50     continue
 60   continue

      end
\end{verbatim}

\subsection{Compressed Jacobian} \label{app:expl:cpr}

\subsubsection{explosion.adf} This is the same as in the dense case.

\subsubsection{explosion.driver.compressed.f} 

\begin{verbatim}
      program main
      implicit none
C 
C     Example: Explosion Equation
C     Driver for computing compressed Jacobian
C
      integer dim, parmax, n
      parameter (dim=7, parmax=2, n=5)  
      integer i,j
C     independent variables   
      double precision x(dim), prm(parmax)
C     derivative components of independent variables   
      double precision g_x(n,dim)
      double precision g_prm(n,parmax)
C     dependent variables
      double precision f(dim)
C     derivative components of dependent variables
      double precision g_f(n,dim)

C     Initialization of input variables
      x(1) = 1.72  
      x(2) = 3.45  
      x(3) = 4.16  
      x(4) = 4.87  
      x(5) = 4.16  
      x(6) = 3.45 
      x(7) = 1.72 
      prm(1) = 1.3  
      prm(2) = 0.245828 

C     Seeding (CPR)
      do 20 i=1,n
        do 10 j=1,parmax
          g_prm(i,j)=0.0
 10     continue
 20   continue

      g_prm(n-1,1)=1.0
      g_prm(n,2)=1.0

      do 40 i=1,n
        do 30 j=1,dim
          g_x(i,j)=0.0
 30     continue
 40   continue

      g_x(1,1)=1.0
      g_x(1,4)=1.0
      g_x(1,7)=1.0
      g_x(2,2)=1.0
      g_x(2,5)=1.0
      g_x(3,3)=1.0
      g_x(3,6)=1.0

C     call differentiated subroutine
      call g_expl(n,dim,parmax,x,g_x,n,prm,g_prm,n,f,g_f,n);
     
C     print compressed Jacobian
      do 60 i=1,dim
        do 50 j=1,n
          print*, "f'(", i, ",", j, ")=",g_f(j,i)
 50     continue
 60   continue

      end
\end{verbatim}

\subsection{Sparse Jacobian} \label{app:sparslinc}

\subsubsection{explosion.sparse.adf} 

In order to make ADIFOR generate derivative code that can use the sparse
forward mode provided by SparsLinC, the parameter {\tt AD\_FLAVOR}
must be set to {\tt sparse}.

\begin{verbatim}
AD_PROG = explosion.cmp
AD_FLAVOR = sparse
AD_TOP  = expl
AD_PMAX = 9
AD_IVARS= x 
AD_DVARS= f
\end{verbatim}

\subsubsection{explosion.driver.sparse.f} 

\begin{verbatim}
      program main
      implicit none
C 
C     Example: Explosion Equation
C     Driver for computing sparse Jacobian using SparsLinC
C
      integer dim, parmax, n
      parameter (dim=7, parmax=2, n=9)  
      integer i,j
C     independent variables   
      double precision x(dim) 
C     passive inputs
      double precision prm(parmax)
C     pointers to sparse derivative components of 
C     independent variables   
      integer g_x(dim)
C     dependent variables
      double precision f(dim)
C     pointers to sparse derivative components of 
C     dependent variables   
      integer g_f(dim)

C     (index, value) pairs
      integer indexes(dim) 
      double precision values(dim)

C     values used in extraction routine
      integer outlen, info



C     Initialization of input variables
      x(1) = 1.72  
      x(2) = 3.45  
      x(3) = 4.16  
      x(4) = 4.87  
      x(5) = 4.16  
      x(6) = 3.45 
      x(7) = 1.72 
      prm(1) = 1.3  
      prm(2) = 0.245828 

C     Initialization of sparse data structures
      call XSPINI


C     Seeding (sparse identity)
      do 10 i=1,dim
        g_x(i)=0
        g_f(i)=0
        call DSPSD(g_x(i),i,1.d0,1)
 10   continue

C     Call differentiated subroutine

      call g_expl(dim,parmax,x,g_x,prm,f,g_f)

C     Extract derivative components
      do 30 i=1,dim
        call DSPXSQ(indexes,values,dim,g_f(i),outlen,info)
        if (info.eq.0) then
          do 20 j=1,outlen
            print*, "indexes(", i, ",", j, ")=", indexes(j)
            print*, "values(", i, ",", j, ")=", values(j)
 20       continue
        endif
 30   continue

      end
\end{verbatim}

\section{ADIC 1.1}
\label{app:ADIC}

The following files can be downloaded from
\begin{center}
{\tt http://www-unix.mcs.anl.gov/\~\!naumann/ad\_tools.html}
\end{center}

\begin{verbatim}
explosion.c  
explosion.driver.c  
explosion.init  
makefile
\end{verbatim}

\subsection{makefile}

\begin{verbatim}
AD_INC = -I$(ADIC)/include -I.
AD_LIB = -L$(ADIC)/lib/$(ADIC_ARCH)

all:
	adiC -d gradient -i explosion.init
	g++ -g -o explosion.ad $(AD_INC) $(AD_LIB) \
	explosion.ad.c explosion.driver.c \
	-lADIntrinsics-C -laif_grad -lm

clean:
	rm explosion.ad* 
	rm ad_deriv.h
	rm *\~
\end{verbatim}

The environment variable {\tt ADIC} must be set to the directory
in which ADIC has been installed.

\subsection{explosion.init} 

\begin{verbatim}
[SOURCE_FILES]
  explosion.c

[gradient]
  GRAD_MAX=7
\end{verbatim}

The script file specifies the input files (only one in this case)
and a variety of other parameters that can be looked up
in \cite{ADICman}. 
The definition of
{\tt GRAD\_MAX}, the length of the derivative components,
is important.
Its default value is 5, which would cause trouble in our case where
{\tt dim}=$7 > 5.$ Refer to \cite{ADICman} for other ways to set
the value of {\tt GRAD\_MAX}.

\subsection{explosion.driver.c} 

\begin{verbatim}
#include "ad_deriv.h"
#include <stdio.h>
#include <iostream>
#include <stdlib.h>           

extern void ad_explosion(int, DERIV_TYPE*, DERIV_TYPE*, DERIV_TYPE*);

void main()
{
  int i,j;
  int dim=7;                   
  int parmax=2;               
  // independent variables
  DERIV_TYPE *x = new DERIV_TYPE[dim];
  DERIV_TYPE *prm = new DERIV_TYPE[parmax];
  // independent variables
  DERIV_TYPE *F = new DERIV_TYPE[dim]; 
  // independent variables
  InactiveDouble *jac = new InactiveDouble[dim];

  ad_AD_Init(dim);

  ad_AD_SetIndepArray(x,dim);
  ad_AD_SetIndepDone();

  DERIV_val(x[0]) = 1.72; 
  DERIV_val(x[1]) = 3.45; 
  DERIV_val(x[2]) = 4.16; 
  DERIV_val(x[3]) = 4.87; 
  DERIV_val(x[4]) = 4.16; 
  DERIV_val(x[5]) = 3.45;
  DERIV_val(x[6]) = 1.72;
  DERIV_val(prm[0]) = 1.3; 
  DERIV_val(prm[1]) = 0.245828;

  ad_explosion(dim,x,prm,F); 

  for (i=0;i<dim;i++) {
    ad_AD_ExtractGrad(jac,F[i]);
    for (j=0;j<dim;j++) 
      cout << "f'[" << i+1 << "," << j+1 << "]=" << jac[j] << endl;
  }

  ad_AD_Final();
}
\end{verbatim}

\section{ADOL-C}
\label{app:adolc}

\subsection{Modifications of Original Source Code}
\begin{verbatim}
#include <adolc.h>
#include <SPARSE/sparse.h>
#include <stdio.h>
#include <iostream>
#include <stdlib.h>           

extern void explosion_ad(int, adouble*, adouble*, adouble*);

void main()
{
  int i,j;
  int dim=7;                   
  int parmax=2;               
  int tag = 1;


  // independent variables (passiv)
  double *v = new double[dim+parmax];
  // dependent variables (passiv)
  double *Fp = new double[dim]; 


  // independent variables (active)
  adouble *x = new adouble[dim];
  adouble *prm = new adouble[parmax];
  // dependent variables (active)
  adouble *F = new adouble[dim]; 

 
  v[0] = 1.72; 
  v[1] = 3.45; 
  v[2] = 4.16; 
  v[3] = 4.87; 
  v[4] = 4.16; 
  v[5] = 3.45;
  v[6] = 1.72;
  v[7] = 1.3; 
  v[8] = 0.245828;


  trace_on(tag);
    for(j=0;j<dim;j++)
      x[j] <<= v[j];
    for(j=0;j<parmax;j++)
      prm[j] <<= v[j];

    explosion_ad(dim,x,prm,F); 

    for(j=0;j<dim;j++)
      F[j] >>= Fp[j];  
  trace_off();  

}
\end{verbatim}
\subsection{Computation of Sparsity Pattern}
\begin{verbatim}
  // Sparsity pattern declarations
  int option[3];        
  unsigned int** Jsp = new unsigned int*[dim];

  for(j=0;j<dim;j++)
    Jsp[j] = new unsigned int[dim+parmax];
  option[0] = 0; // automatic detection for AD mode
  option[1] = 0; // save propagation of bit-pattern
  option[2] = 0; // no output

  ...

  // Sparsety pattern computation after trace_off

  jac_pat(tag,dim,dim+parmax,v,NULL,NULL,Jsp,option);
\end{verbatim} 
\subsection{Jacobian Calculation Using Low-Level Routines}
\begin{verbatim}
  // Calculate Jacobian using forward mode driver
  double** X = myalloc(dim+parmax,dim+parmax);         
  // Calculate Jacobian using reverse mode driver
  double** U = myalloc(dim,dim);   

  ...

  // Use low level routines to compute Jacobian
  // forward:

  for (j=0;j<dim+parmax;j++) 
  {
   for (i=0;i<dim+parmax;i++) 
     X[j][i] = 0.0;
   X[j][j] = 1.0;
  }

  // first order vector mode forward
  // ^     ^     ^           ^^^^^^^ = fov_forward
  fov_forward(tag,dim,dim+parmax,dim+parmax,v,X,Fp,J);

  // reverse
  for (j=0;j<dim;j++) 
  {
   for (i=0;i<dim;i++) 
     U[j][i] = 0.0;
   U[j][j] = 1.0;
  }

  // prepare reverse sweep with appropriate forward sweep
  zos_forward(tag,dim,dim+parmax,1,v,Fp);

  // first order vector mode reverse
  // ^     ^     ^           ^^^^^^^ = fov_reverse
  fov_reverse(tag,dim,dim+parmax,dim,U,J); 

\end{verbatim} 
\section{TAPENADE} 
\label{app:TAPENADE}

The following files can be downloaded from
\begin{center}
{\tt http://www-unix.mcs.anl.gov/\~\!naumann/ad\_tools.html}
\end{center}

\begin{verbatim}
explosion.f  
explosion.driver.f  
makefile
\end{verbatim}

\subsection{makefile}
\label{app:TAPENADEmake}
\begin{verbatim}
ADSTACK = $(HOME)/ADTOOLS/TAPENADE/tapenade1.0/stack/adStack.o

all:
	tapenade -head expl -vars "x prm" -cl explosion.f
	g77 -o explosion.ad explcl.f explosion.driver.f \
	$(ADSTACK)

clean:
	rm explosion.ad
	rm explcl.f
	rm -fr diffgen
	rm *\~
\end{verbatim}

\subsection{explosion.driver.f} 
\label{app:TAPENADEdriver}

\begin{verbatim}
      program main
      implicit none
C 
C     Example: Explosion Equation
C
C     Driver for computing Jacobian using TAPENADE's 
C     Reverse Mode
C

      integer dim, parmax, n
      parameter (dim=7, parmax=2, n=9)  
      integer i,j
C     independent variables   
      double precision x(dim), prm(parmax)
C     derivative components of independent variables   
      double precision g_x(dim)
      double precision g_prm(parmax)
C     dependent variables
      double precision f(dim)
C     derivative components of dependent variables
      double precision g_f(dim)
C     the whole Jacobian matrix
      double precision jac(dim,n)

C     Initialization of input variables
      x(1) = 1.72  
      x(2) = 3.45  
      x(3) = 4.16  
      x(4) = 4.87  
      x(5) = 4.16  
      x(6) = 3.45 
      x(7) = 1.72 

      prm(1) = 1.3  
      prm(2) = 0.245828 

C     Compute Jacobian as sequence of Jacobian transposed
C     times vector products
      do 40 i=1,dim
        do 10 j=1,dim
          g_f(j)=0.d0
          g_x(j)=0.d0
 10     continue
        g_f(i)=1.d0
        g_prm(1)=0.d0
        g_prm(2)=0.d0
        call explcl(dim,parmax,x,g_x,prm,g_prm,f,g_f);
        do 20 j=1,dim
          jac(i,j)=g_x(j)
 20     continue
        do 30 j=1,parmax
          jac(i,j+dim)=g_prm(j)
 30     continue
 40   continue

C     print Jacobian
      do 60 i=1,dim
        do 50 j=1,n
          print*, "f'(", i, ",", j, ")=",jac(i,j)
 50     continue
 60   continue

      end
\end{verbatim}

\end{document}